\documentclass[%
reprint, superscriptaddress, amsmath,amssymb, aps,
 pre,  %
]{revtex4-2}

\usepackage{graphicx}
\usepackage{dcolumn}
\usepackage{braket}
\usepackage{amsmath}
\usepackage{xcolor}
\usepackage{comment}
\usepackage{dsfont}
\usepackage{bbm}
\usepackage{bm}
\usepackage[normalem]{ulem}   %
\usepackage[colorlinks=true,allcolors=blue]{hyperref}

\begin{document}

\preprint{APS/123-QED}

\title{Dissipation-Reliability Tradeoff for Stochastic CMOS Bits in Series}

\author{Cathryn P. Murphy}
 \affiliation{Department of Chemistry, Northwestern University, 2145 Sheridan Road, Evanston, Illinois 60208, USA}
\author{Schuyler B. Nicholson}
\affiliation{Santa Fe Institute, 1399 Hyde Park Road, Santa Fe, New Mexico 87501, USA}
\affiliation{Department of Chemistry, Northwestern University, 2145 Sheridan Road, Evanston, Illinois 60208, USA}

\author{Nahuel Freitas}
\affiliation{Extropic Corporation, Cambridge, Massachusetts, USA}
\affiliation{Departamento de F\'isica, FCEyN, UBA, Pabell\'on 1, Ciudad Universitaria, 1428 Buenos Aires, Argentina}

\author{Emanuele Penocchio}
 \email{emanuele.penocchio@uni-mainz.de}
\affiliation{Department of Chemistry, Johannes Gutenberg University Mainz, Duesbergweg 10--14, 55128 Mainz, Germany}
\affiliation{Department of Chemistry, Northwestern University, 2145 Sheridan Road, Evanston, Illinois 60208, USA}

 \author{Todd R. Gingrich}
 \email{todd.gingrich@northwestern.edu}
\affiliation{Department of Chemistry, Northwestern University, 2145 Sheridan Road, Evanston, Illinois 60208, USA}
\affiliation{Department of Engineering Sciences and Applied Mathematics, Northwestern University, 2145 Sheridan Road, Evanston, Illinois 60208, USA}

\date{\today}

\begin{abstract}
Physical instantiations of a bit of information are subject to thermal noise that can trigger unintended bit-flip errors.
Bits implemented with CMOS technology typically operate in regimes that reliably suppress these errors with a large supply voltage, but miniaturization and circuit design for implantable biomedical devices motivate error suppression via alternative low-voltage strategies.
We present and analyze an error-suppression technique that involves coupling multiple CMOS units into chains, introducing a natural error correction arising from inter-unit correlations.
Using tensor networks to numerically solve a stochastic master equation for the CMOS chain, we quantify the dissipation-reliability tradeoff across system sizes that would be intractable with conventional sparse-matrix methods.
The calculations show that the typical time for bit-flip errors grows exponentially in the square of the supply voltage but subexponentially with the chain length, a scaling we attribute to a spatially localized, domain-wall-mediated flip rather than a coherent one.
While a CMOS chain adds stability compared to a single CMOS unit for a fixed low supply voltage, increasing the supply voltage is a lower-dissipation route to equivalent stability.
\end{abstract}

\maketitle

\section{Introduction}

Classical computers rely on robust maintenance of logical bits, and any physical instantiation of those bits is subject to thermal noise~\cite{Wang06}.
That noise can stochastically trigger unintended bit-flip errors~\cite{Natori88,Rezaei20}, and it is important to be able to suppress the errors.
A common strategy is implemented in CMOS (complementary metal-oxide-semiconductor) devices, which involve complementary transistors powered by a supply voltage.
When the supply increases, errors decrease, but at the expense of dissipating more energy.
Engineering high-accuracy and low-dissipation devices has been a theoretical and practical goal for many decades~\cite{Landauer61,Bryant01,Bol09,Hanson08,Kim11,Aifer24,Riechers20}.

Miniaturization and circuit design for implantable biomedical devices \cite{Rivnay13,Sun18,Jia21} both require low-power computing strategies, so we consider the dissipation-reliability balance when the supply voltage cannot simply be increased.
The limitation on the supply voltage motivates us to consider an alternative way to mitigate errors: by building multiple error-prone CMOS units into a larger bistable system that encodes a single logical bit.
We show that as the number of units in series increases, the bit grows more stable.
Correlations between the units introduce a natural error correction that slows the rate of stochastic bit flips.
The stability benefit is countered by the increased thermodynamic cost; each unit dissipates energy.
While linear scaling of dissipation is easy to justify in terms of contributions from each unit, those units are strongly interacting.
Errors require a rare event in one unit to trigger rare events in the others so the entire circuit can flip in a coordinated manner.

To demonstrate how that correlation effect emerges from essential physics, we build up to the collective dynamics by starting with a model that explicitly resolves the dynamics and thermodynamics at the single electron level.
Each electron hop triggers a transition to a new discrete state of the circuit.
One can conceptually picture the circuit as if its state were confined to one basin of a double-well potential, occasionally flying over a barrier due to some Gaussian noise from the environment.
That picture captures the essential bistability, but fails to provide a thermodynamically consistent connection between the energetics of the circuit and the thermal fluctuations \cite{Weiss98,Sarpeshkar93,Wyatt99,Hanggi88,Freitas22}.
Previous work has established a stochastic model for a CMOS circuit that describes the dynamics of the elementary steps and accounts for the thermodynamic cost~\cite{Freitas21,Gao21,Gu19}, validated by agreement with experiments~\cite{Freitas26}.
The master-equation model was numerically solved, giving access to both error and dissipation rates~\cite{Freitas22, Gao21}.
Extending that approach to larger circuits is tempting to do via coarser descriptions\textemdash replacing discrete electron hops with continuous voltage noise\textemdash but doing so severs the thermodynamically consistent link between circuit energetics and thermal fluctuations that local detailed balance enforces at each elementary step.
One is therefore caught between two undesirable options: retain microscopic discreteness and face a state space too large for conventional methods, or adopt a tractable coarse-grained description and sacrifice thermodynamic fidelity.
Tensor network methods resolve this tension by efficiently representing the high-dimensional probability distribution without explicitly enumerating the exponentially large state space.

\begin{figure*}
\includegraphics[width=\textwidth]{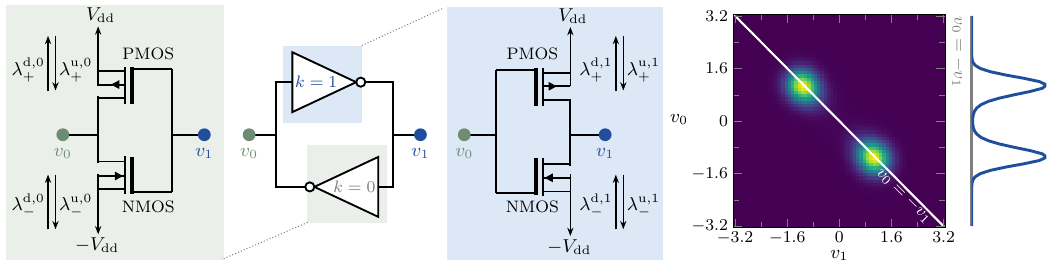}
\caption{\label{fig:fig1} The minimal instantiation of a bit with CMOS technology is a single unit with two nodes at voltages $v_0$ and $v_1$.
These nodes are coupled by two NOT gates, each consisting of an n-type ($-$) and p-type ($+$) transistor.
In all four transistors, electrons hop in both directions between ideal thermodynamic reservoirs ($-V_{\rm dd}$ and $V_{\rm dd}$) and the fluctuating voltage nodes.
Rates for those hops, $\lambda$, depend on $\mathbf{v}$, with three indices indicating the NOT gate, the n/p transistor, and the direction of the hop.
The coupled nodes yield a bistable steady state (right), with each peak corresponding to a metastable bit-state.}
\end{figure*}

Here, we introduce tensor network (TN) methods \cite{Evenbly11,Kazeev14,Vo17,Nicholson23,Zima25} combined with a judicious basis set expansion to stochastic electronic circuits, numerically solving for the steady-state thermodynamics (dissipation rate) and the relaxation timescales (error rate) for bistable coupled circuits.
The main results are threefold.
First, we demonstrate that a higher accuracy can be achieved either by increasing the supply voltage or by daisy chaining together units, and that the former strategy results in less dissipation while the latter can be useful for low-voltage computing.
Second, we find that $\ln \langle \tau_{\rm err} \rangle$ grows \emph{sublinearly} with the chain length $L$, qualitatively distinct from the linear-in-$L$ growth of the analogous reaction-diffusion chain~\cite{Nicholson23}.
We argue in Sec.~\ref{sec:results_L} that this sublinear scaling, together with an isolated slow relaxation mode, supports domain-wall nucleation, not a coherent flip of the entire chain, as the rare event behind each bit-flip error.
Third, we establish tensor network methods for stochastic circuit models that faithfully resolve both rare events and thermodynamics, preserving the discrete thermal noise from which thermodynamic consistency emerges even for systems exceeding $10^{14}$ microstates.
We propose that our framework be applied to other computing hardware in which thermal noise shapes the behavior, whether as a bug to suppress or, as in thermodynamic computing~\cite{Freitas26,Chowdhury23,TCconference}, a feature to exploit.
We have focused on CMOS technology, distinguished by its compatibility with the existing semiconductor manufacturing infrastructure and by the fact that the same circuit element provides both the stochastic source and the digital interface, but CMOS is, of course, not the only platform for a bit.
Magnetic tunnel junctions with tunable thermal energy barriers~\cite{Bhatti17,Borders19}, optical systems~\cite{BP24}, and Josephson-junction circuits~\cite{Elmitwalli23} all realize probabilistic bits (p-bits), and each platform can be expected to have its own version of a dissipation-reliability tradeoff.

The paper is structured as follows.
Section~\ref{sec:single} develops the dynamics and thermodynamics of a single CMOS unit, along with the assumptions that define the model's regime of validity, and Sec.~\ref{sec:chain} couples $L$ such units into a chain.
Section~\ref{sec:tn} details the tensor network methodology used to solve the resulting many-body master equation, and Sec.~\ref{sec:results} presents the main numerical results: the voltage and length scalings of the mean bit-flip time, and the resulting dissipation-reliability tradeoff, before we conclude in Sec.~\ref{sec:conclusion}.

\section{Dynamics and Thermodynamics of a Single CMOS Unit}
\label{sec:single}

\subsection{A Single CMOS Unit and Its Hop Rates}
\label{sec:unit}
A single CMOS unit is typically built from two NOT gates, each consisting of an n-type and p-type metal oxide semiconductor field effect transistor (MOSFET) through which electrons can hop bidirectionally between a source at voltage $-V_{\rm dd}$ and a drain at voltage $V_{\rm dd}$.
The permitted electron hops associated with each transistor in the single unit are shown in Figure~\ref{fig:fig1}.
The unit's state is given by the vector $\mathbf{v} = (v_0, v_1)$, with voltages $v_0$ and $v_1$ measured at nodes on opposite sides of the unit.
The hop rates depend on the state of the circuit through $\mathbf{v}$; the characteristics of the device through a threshold voltage $V_{\rm th}$, specific current $I_0$, and subthreshold slope factor $n$; and the temperature $T$ through a thermal voltage $V_T = k_{\rm B} T / e$, where $k_{\rm B}$ is Boltzmann constant and $e$ the charge of an electron.
The subthreshold slope factor $n \geq 1$ quantifies the deviation of the gate-coupling efficiency from the ideal thermionic limit, with $n = 1$ corresponding to perfect coupling and typical fabricated MOSFETs having $n$ in the range $1.0$--$1.5$~\cite{Nowak02}.

Throughout the paper we report voltages in units of the thermal voltage $V_T$, which, at $T = 300$~K, is about $25.85~\mathrm{mV}$.
Though $V_{\rm th}$ and $I_0$ appear in the hopping rates, we absorb them into a constant $\tau_0=\frac{e}{I_0}\exp({V_{\rm th}/nV_T})$, which scales the forward and backwards hopping rates equally.
We use $\tau_0$ as the natural unit to report timescales.
Following Ref.~\cite{Freitas21}, $V_{\rm th}$ and $I_0$ enter the model only through $\tau_0$, so we quote no values for them.
The constant does carry a direct physical reading.
In the bistable phase the mean deterministic current through a transistor is $\left<I\right> = e/\tau_0$~\cite{Freitas22}, so $\tau_0$ is the average time to move a single electron through a transistor.

With two orientations of NOT gates and two transistors per gate, there are four junctions across which electrons hop.
Each elementary hop also includes a direction, either ``uphill'' or ``downhill'', referring to the direction of the mean flow of electrons (from source to drain).
The hops are frequently modeled as Poissonian transitions which alter the voltage at one of the two nodes by a discrete amount $v_e = e / C$, where $C$ is the capacitance of the transistor.
Modeling the hops as Poissonian relies on a quasi-steady-state assumption that we state, together with the model's other assumptions, in Sec.~\ref{sec:assumptions}.
We write the rate of an uphill hop as
\begin{equation}
  \lambda_{\pm}^{{\rm u}, k}(\mathbf{v}) = \tau_0^{-1}\exp \left(\frac{V_{dd} \mp \delta_{k, 1} v_0 \mp \delta_{k, 0} v_1}{nV_T}\right),
  \label{eq:eq1}
\end{equation}
where the $\pm$ subscript differentiates between hops in the n-type (-) and p-type (+) MOSFETs and the $k$ index selects for the bottom NOT gate ($k = 0$, shaded gray) or the top ($k = 1$, shaded blue) in Fig.~\ref{fig:fig1}.
Hops in NOT gate $k$ cause $v_k$ to change while the voltage of the other node stays fixed.
The Kronecker $\delta$ functions of Eq.~\eqref{eq:eq1} imply that uphill steps that change each node's voltage occur with a rate that depends only on the value of the other node's voltage.
The rates of downhill hops take a similar form, but depend on both $v_0$ and $v_{1}$:
\begin{equation}
  \lambda_{\pm}^{{\rm d}, k}(\mathbf{v}) = \lambda^{{\rm u},k}_{\pm}(\mathbf{v}) \exp \left(\frac{-V_{dd}-\frac{1}{2}v_e \pm \delta_{k, 0} v_0 \pm \delta_{k, 1} v_1}{V_T}\right).
  \label{eq:eq2}
\end{equation}
For every electron that hops downhill at a p-type MOSFET or uphill at an n-type, the altered voltage ($v_0$ or $v_{1}$) goes down by the discrete step $v_e$.
Similarly, jumps in the opposite direction increase voltages by that same $v_e$.

\subsection{Stochastic Dynamics and Thermodynamics}
\label{sec:dynamics}
The discrete hops cause the voltage to fluctuate according to a master equation
\begin{equation}
\frac{\partial \vec{P}}{\partial t} = \mathbb{W} \vec{P},
\end{equation}
where $\vec{P}$ is a vector of probabilities of each microstate, specified by the voltage values at both nodes $0$ and $1$.
Because each of these voltages can take a discrete number of possible values, the operator $\mathbb{W}$ is the usual transition rate matrix with off-diagonal elements expressed in terms of $\lambda_{\pm}^{{\rm u}, k}$ and $\lambda_{\pm}^{{\rm d}, k}$ and diagonal elements ensuring the conservation of probability.
The model is therefore a continuous-time Markov jump process; no time discretization or fixed update order is imposed.
Any forward-time stochastic trajectory can be generated, e.g., by the Gillespie algorithm~\cite{Gillespie76}, but our numerical results instead come from a direct eigensolution of $\mathbb{W}$, as described in Sec.~\ref{sec:tn}.
The stochastic dynamics samples values of $v_0$ and $v_{1}$ from the non-equilibrium steady-state distribution $P_{\rm ss}(\mathbf{v})$ that solves $\mathbb{W} \vec{P} = 0$.
The symmetry of the gates that make up the CMOS circuit furthermore ensures a symmetry in the probabilities, $P_{\rm ss}(\mathbf{v}) = P_{\rm ss}(-\mathbf{v})$, that yields a bistability, shown in Fig.~\ref{fig:fig1}.
A voltage drop across the circuit is equally likely to be positive as negative, allowing the CMOS circuit to be thought of as a physical instantiation of a logical bit.
The first excited state eigenvalue and eigenvector of $\mathbb{W}$ give the timescale and mode of relaxation between the two states of the bit \cite{Roux22,Zima25}.

It is also interesting to consider flows of energy that accompany the stochastic dynamics.
Elementary uphill and downhill hops move electrons between ideal thermodynamic reservoirs held at particular voltages and temperature $T$.
Each such hop is microscopically reversible: the reverse hop occurs as well.
The ratio of forward and reverse rates is fixed by the requirement that the circuit, were it coupled to a single thermal bath with no driving, relax to equilibrium with the correct Boltzmann statistics.
That constraint on the rates of Eqs.~\eqref{eq:eq1}--\eqref{eq:eq2} is the principle of local detailed balance~\cite{Seifert05,Seifert12,Esposito10,Falasco21}.
Concretely, local detailed balance imposes
\begin{equation}
  -k_{\rm B} T \log\frac{\lambda_{\pm}^{{\rm u},k}(\mathbf{v})}{\lambda_{\pm}^{{\rm d},k}(\mathbf{v} \pm v_e(\delta_{k, 0}, \delta_{k, 1}))} = \delta Q_{\pm}(v_k),
  \label{eq:ldb}
\end{equation}
where
\begin{equation}
  \delta Q_{\pm}(v_k) = e\left(-V_{dd} +\frac{1}{2}v_e \pm v_k\right)
  \label{eq:deltaQ}
\end{equation}
is the energy flow from bath to system when moving a single electron between node $k$ and a thermodynamic reservoir (source/drain) in the uphill $v_k \to v_k \pm v_e$ direction~\cite{Freitas21}.
This local detailed balance condition is what ties the circuit energetics directly to the thermal fluctuations, and it is the property we must preserve when extending to larger circuits.
In the steady state, the net rate of these uphill steps is the particle current
\begin{equation}
  \begin{split}
    j_{\pm}^k(\mathbf{v}) =& \lambda^{{\rm u},k}_{\pm}(\mathbf{v}) P_{\rm ss}(\mathbf{v}) \\
    &- \lambda^{{\rm d},k}_{\pm}(\mathbf{v} \pm v_e(\delta_{k, 0}, \delta_{k, 1})) P_{\rm ss} (\mathbf{v} \pm v_e(\delta_{k, 0}, \delta_{k, 1})),
    \label{eq:currents}
  \end{split}
\end{equation}
where the $\pm$ again distinguishes between the current through n- and p-type transistors and $k$ selects the NOT gate's orientation.
The average dissipation rate for the entire CMOS circuit sums over contributions from each transistor for each microstate $\mathbf{v}$:
\begin{equation}
  \dot{Q} = \sum_{v_0,v_1}\sum_{\alpha = \pm} \sum_{k} \delta Q_\alpha(v_k) j_{\alpha}^k(\mathbf{v}).
  \label{eq:dissipation}
\end{equation}

\subsection{Assumptions, Parameters, and Regime of Validity}
\label{sec:assumptions}

The model just described rests on a handful of assumptions, which we collect here along with the parameter values used throughout the paper (Table~\ref{tab:params}).
The model was developed independently by two groups~\cite{Freitas21,Gao21} and has since been validated against experiment~\cite{Freitas26}.
Our aim is therefore not to justify the model anew but to state the assumptions that set its regime of validity.

The first assumption is the quasi-steady-state treatment of each hop.
Modeling the hops as Poissonian requires that the relaxation of the channel charge distribution and gate dielectric polarization be fast compared with the inter-hop interval, so that rates depend only on the instantaneous node voltages and not on the history of recent hops.
This assumption is standard for the bias regimes and device dimensions of interest~\cite{Freitas21,Gao21}.

The second assumption is the semiclassical treatment of transport.
The Poissonian-hop framework is appropriate for device length scales above the few-nm regime, where quantum confinement, source-to-drain tunneling, and ballistic transport become significant.
For sub-10-nm nodes, the rates of Eqs.~\eqref{eq:eq1}--\eqref{eq:eq2} would need to be replaced by rates derived from quantum-corrected drift-diffusion, nonequilibrium Green's functions, or similar transport models.
Only the rate functions\textemdash which encode the local detailed balance of the semiclassical model\textemdash would need modification; the master-equation structure and the numerical machinery of Sec.~\ref{sec:tn} carry over once the revised rates are in hand.
The revision is therefore less drastic than it sounds.

The third assumption concerns the noise.
The model includes only the thermal (shot-like) noise of individual electron hops, which is delta-correlated in time, so one hop says nothing about when the next will arrive.
Real devices carry additional noise sources with nonzero correlation times: generation-recombination noise from donor and acceptor states, $1/f$ (flicker) noise~\cite{Pierce56}, and random telegraph signal noise from individual traps~\cite{Luo15}.
Enlarging the state space with auxiliary trap-occupation variables renders the augmented dynamics Markovian again, and parasitic capacitances could similarly enter as site-dependent step sizes $v_e$.
In both cases the price is a larger physical dimension per site.
Paying it is taxing rather than onerous, at least until that dimension grows large, and the basis expansion of Sec.~\ref{sec:basis_funcs} pushes back the point at which it does.

Finally, the parameters.
Table~\ref{tab:params} lists the values used in the calculations of Sec.~\ref{sec:results}.
The capacitance fixes the elementary voltage step $v_e = e/C$; modern transistors can be fabricated with capacitance values as low as $\sim50$~aF~\cite{Zheng16}, corresponding to a value of $v_e/V_T\approx 0.1$ at room temperature.
Note that operating devices sit far above the supply voltages studied here.
Conventional room-temperature logic runs at $V_{\rm dd}\sim0.7$--$1$~V, i.e., $V_{\rm dd}/V_T\gtrsim30$, and even the low-voltage bioelectronics that motivate low-power alternatives~\cite{Rivnay13,Sun18,Jia21} operate at hundreds of millivolts, an order of magnitude above the near-threshold window $V_{\rm dd}/V_T\in[1.1,1.4]$ that we examine.
The window is nevertheless a natural one.
For an ideal device the minimum supply voltage at which a CMOS inverter can operate is $2\ln 2\,V_T \approx 36$~mV~\cite{Swanson72,MeindlDavis00}, and since our supply spans $-V_{\rm dd}$ to $V_{\rm dd}$ that limit is reached at $V_{\rm dd}/V_T = \ln 2$, exactly where the chain first becomes bistable.
We work between $1.6$ and $2$ times that minimum.
Closer to the onset the bit is only marginally stable, and further above it spontaneous flips become too rare to resolve.
The scaling laws are what transfer.
Extrapolated to device-relevant supply voltages, the scalings established in Sec.~\ref{sec:results} predict the stability of bits in regimes where flips, though inaccessible to direct computation, remain relevant to engineering.

\begin{table}[h]
\caption{Parameters used in this work at $T = 300$~K.}
\label{tab:params}
\begin{ruledtabular}
\begin{tabular}{lc}
Parameter & Value \\
\hline
$C$ & $\sim 50$~aF \\
$V_T$ & $\sim 25.85$~mV \\
$V_{\rm dd}/V_T$ & $[1.1$:$1.4]$ \\
$v_e/V_T$ & $0.1$ \\
$n$ & $1.0$ \\
\end{tabular}
\end{ruledtabular}
\end{table}

\section{Coupling CMOS Units to Make a Collective Bit}
\label{sec:chain}

Having established the dynamics and thermodynamics of a single CMOS unit, we now build a larger circuit from $L$ units, connected in series.
The state $\mathbf{v}$ of the CMOS chain is characterized by a higher dimensional vector $\mathbf{v} = (v_0, v_1, \ldots, v_L)$ whose components $v_i$ indicate the voltage of each node between (and at the boundaries of) the units.
The $i$-th unit ($i = 0, 1, \ldots, L-1$) is bracketed by nodes $i$ and $i+1$, so that consecutive units share a node and each interior node is influenced by hops in two adjacent units.
The rate of elementary hops across a MOSFET in unit $i = 0, 1, \hdots L-1$ is again given by expressions of the form given in Eqs.~\ref{eq:eq1} and \ref{eq:eq2} except that elementary hops within unit $i$ now depend on $v_i$ and $v_{i+1}$.
Explicitly, with the same conventions as for the single-unit case, the uphill rates in unit $i$ are
\begin{align}
  \lambda_{\pm,i}^{{\rm u},{\rm b}}(\mathbf{v}) &=\tau_0^{-1} \exp\!\left(\frac{V_{\rm dd}  \mp v_{i+1}}{nV_T}\right), \\
  \lambda_{\pm,i}^{{\rm u},{\rm t}}(\mathbf{v}) &= \tau_0^{-1}\exp\!\left(\frac{V_{\rm dd} \mp v_{i}}{nV_T}\right),
  \label{eq:rates_chain_uphill}
\end{align}
where the ``b''/``t'' superscript distinguishes the bottom and top NOT gates of unit $i$, and the downhill rates follow from local detailed balance applied to the corresponding hops, in direct analogy with Eq.~\eqref{eq:eq2}.
As before, each discrete hop changes one of these voltages by $v_e$, and hops in different units are independent Poisson processes.

Provided that the supply voltage is large enough to support bistability ($V_{\rm dd}/V_T \geq \ln 2$ for $n=1$), voltages at neighboring nodes become strongly anti-correlated with each other.
$P_{\rm ss}(\mathbf{v})$ for a chain with an odd number of $L$ units is therefore dominated by one of two symmetrically equivalent macrostates, each with large voltages of alternating sign.
The transition from one macrostate to the other is a rare event that requires all nodes to flip their voltage, so the CMOS chain acts as a single collective bit.
The characteristic timescale for that bit flip, $\left<\tau_{\rm err}\right>$, depends on the number of units in the chain, with longer chains suppressing the rate of spontaneous bit flips.
An alternative strategy to increase $\left<\tau_{\rm err}\right>$ is to increase the supply voltage $V_{\rm dd}$.
Both strategies to enhance bit stability come with a dissipative cost.

\section{Tensor Network Methodology}
\label{sec:tn}

\begin{figure}
\includegraphics[width=\columnwidth]{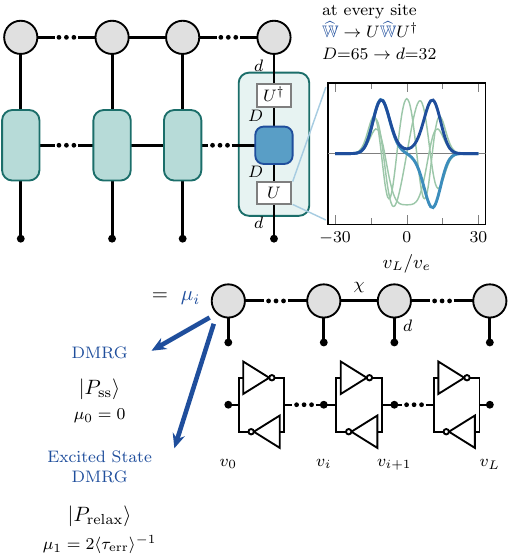}
\caption{\label{fig:fig2} The eigenvalue problem for the rate operator, Eq.~\eqref{eq:rateoperator}, drawn as a tensor network.
Circles are MPS tensors and rounded boxes are MPO tensors, and a line joining two of them is an index summed over.
At the highlighted site the reduced MPO, in teal, is opened up to reveal the full rate operator in blue, whose physical indices run over the $D = 65$ voltage states of a node, sandwiched between the isometry $U$ and its adjoint, which map to and from the $d = 32$ retained basis functions.
Those functions are the rows of $U$, plotted in the inset, where the first two are highlighted as the single unit's steady state and its slowest relaxation mode (Sec.~\ref{sec:basis_funcs}).
The two tensor rows are the two sides of $\hat{\mathbb{W}}\ket{P_i} = \mu_i\ket{P_i}$, and the lower one carries bond dimension $\chi$ and physical dimension $d$.
The CMOS chain beneath shares its abscissae, so each tensor is one node, but the leg above each node runs over basis coefficients rather than over the voltage $v_i$ itself.
The density matrix renormalization group (DMRG) and its excited-state variant return $\ket{P_{\rm ss}}$ and $\ket{P_{\rm relax}}$, the eigenvectors of $\mu_0 = 0$ and of the spectral gap $\mu_1 = 2\langle\tau_{\rm err}\rangle^{-1}$. }
\end{figure}

To analyze tradeoffs, we need a numerical strategy to compute both $\left<\tau_{\rm err}\right>$ and $\dot{Q}$ for CMOS chains.
In principle, the calculations are straightforward.
The top right eigenvector of $\mathbb{W}$ is $P_{\rm ss}(\mathbf{v})$, from which particle currents and dissipation are computed as in Eqs.~\ref{eq:currents} and \ref{eq:dissipation}.
The next eigenvector captures the slowest relaxation mode, with associated eigenvalue $\mu_1$ giving the flipping timescale through: $\langle\tau_{\rm err}\rangle=2 / \mu_1$~\cite{Roux22}.
In practice, calculations of the principal eigenvector and the spectral gap are nontrivial because $L$-unit chains introduce a many-body problem.
Consequently, we adopt a TN approach, computing numerically exact results with enormous state spaces ($65^8\approx3\times10^{14}$ microstates for the $L = 7$ chain at the $\pm32v_e$ cutoff of Sec.~\ref{sec:state_space}) with the density matrix renormalization group (DMRG).

\subsection{State Space Construction}
\label{sec:state_space}
The state of each site corresponds to the voltage of a node, which can only be a discrete multiple of $v_e$.
In practice, the CMOS architecture prevents the magnitudes of the voltages from growing without bound, so it is reasonable to truncate the allowable voltage values at some large positive and negative threshold.
For the calculations we present in Sec.~\ref{sec:results}, it was sufficient to allow voltages to run from $-32 v_e$ to $32 v_e$.
This cutoff is generous.
The steady-state distribution of $v_i$ has typical magnitude of order $V_{\rm dd}/v_e \approx 14$ for the parameters used here, with thermal fluctuations of much smaller amplitude.
We verified empirically that increasing the cutoff to $\pm 40 v_e$ produces no detectable change in the reported observables.
The state of the full circuit is given by the set of states corresponding to each node in the chain.
For a circuit consisting of $L+1$ nodes whose voltages are allowed to range from $-32v_e$ to $32v_e$ the size of the full space is $(32\times2+1)^{(L+1)}$.

Tensor network methods make the problem tractable by never enumerating that full space.
Instead, the joint distribution is approximated to controllable accuracy by a multilinear function, graphically depicted as a partially contracted network of tensors, the elements of which are variational parameters tuned to optimize the approximation.
Those tensors, one per node, carry so-called physical indices which specify the voltage state of the node, a state that varies over the $D = 65$ options.
The cost of this approximation scheme grows with that physical dimension $D$, but much less sharply than the $D^{L+1}$ growth in the number of microstates.

Each single-node tensor is not an independent factor.
A second index, the so-called bond index, links the tensors to their neighbors, and the resulting chain is a matrix product state (MPS)~\cite{Schollwoeck11}.
Imagine cutting the chain between two adjacent nodes.
All correlation between the two halves must pass through the single index spanning that cut, and the size of that index is the bond dimension $\chi$.
Equivalently, the MPS writes the distribution across that cut as a sum of products, each pairing a function of the voltages on the left with a function of those on the right, and $\chi$ is the number of products it retains.
It therefore bounds the Schmidt rank across every cut, and an arbitrary distribution would demand a $\chi$ that grows \emph{exponentially} with $L$.
A small $\chi$ is cheap to manipulate but limits the correlations expressible across each cut, while growing $\chi$ enlarges the set of representable distributions until the representation becomes \emph{exact}.
In other words, the MPS is a controllable approximation rather than a factorization, with $\chi$ the parameter we converge.
Whether a modest $\chi$ suffices depends on how nearly the distribution factorizes.
Nudge the voltage at one node, then at a distant one.
If the effect of the first hardly depends on the second, few products are needed and $\chi$ can stay small.
Though not a priori obvious for all applications, many physical problems can be solved with a severely truncated $\chi$.
Section~\ref{sec:DMRGconvg} provides numerical evidence that this is one such problem.

\subsection{Second-quantized Rate Operator and MPO}
\label{sec:operator_MPO}
To employ the tensor network approach, we first write the rate operator in a second-quantized form.
Each elementary electron hop induces a discrete jump of the voltage at a single node, so we introduce the raising and lowering operators
\begin{equation}
  \hat{x}_i^\pm\ket{\mathbf{v}} = \ket{v_0,\hdots,v_{i}\pm v_e,\hdots,v_L},
\end{equation}
with $\ket{\mathbf{v}}$ the Fock vector for microstate $\mathbf{v}$.
Each voltage jump can be induced by electron hopping events at multiple different transistors, and the rate for each of those hops depends on the voltages.
To compactly express the rates of each event, we introduce, for each CMOS unit $i$, diagonal operators $\hat{t}^{\rm u/d}_{\pm, i}$ and $\hat{b}^{\rm u/d}_{\pm, i}$ giving the hopping rates in the top and bottom NOT gates, respectively.
Mirroring Eqs.~\eqref{eq:eq1} and \eqref{eq:eq2},
\begin{align}
  \nonumber
  \hat{b}_{\pm, i}^{\rm u}\ket{\mathbf{v}} &= \tau_0^{-1}\exp \left(\frac{V_{dd} \mp v_{i+1}}{nV_T}\right) \ket{\mathbf{v}} \\
  \hat{b}_{\pm, i}^{\rm d}\ket{\mathbf{v}} &= \hat{b}_{\pm, i}^{\rm u} \exp \left(\frac{-V_{dd}-\frac{1}{2}v_e \pm v_i}{V_T}\right) \ket{\mathbf{v}}.
  \label{eq:NOT}
\end{align}

Equivalent expressions for $\hat{t}^{\rm u}_{\pm, i}$ and $\hat{t}^{\rm d}_{\pm, i}$ follow by changing $b$'s to $t$'s and exchanging the positions of $v_i$ and $v_{i+1}$ in Eq.~\eqref{eq:NOT}, since the top NOT gate is oriented opposite to the bottom.

The rate operator includes contributions from uphill and downhill hops through all four transistors and sums over all CMOS units:

\begin{widetext}
\begin{equation}
  \hat{\mathbb{W}} =  \sum_{i = 0}^{L-1} \left[\left(\hat{t}^{\rm u}_{+, i} + \hat{t}^{\rm d}_{-, i}\right) \left(\hat{x}_{i+1}^+ - \mathbb{I}_{i+1}\right)  + \left(\hat{t}^{\rm d}_{+, i} + \hat{t}^{\rm u}_{-, i}\right) \left(\hat{x}_{i+1}^- - \mathbb{I}_{i+1}\right) + \left(\hat{b}^{\rm u}_{+, i} + \hat{b}^{\rm d}_{-, i}\right) \left(\hat{x}_{i}^+ - \mathbb{I}_{i}\right)  + \left(\hat{b}^{\rm d}_{+, i} + \hat{b}^{\rm u}_{-, i}\right) \left(\hat{x}_{i}^- - \mathbb{I}_{i}\right)\right].
  \label{eq:rateoperator}
\end{equation}
\end{widetext}

Here, it is understood that $\hat{x}^\pm_i = \mathbb{I}_0 \otimes \hdots \otimes x^\pm_i \otimes \hdots \otimes \mathbb{I}_L$, with implicit identity operators.
The additional explicit identity operators of Eq.~\eqref{eq:rateoperator} generate the negative diagonal elements that ensure probability conservation.
This representation in terms of raising and lowering operators lends itself to an efficient compression in terms of a matrix product operator (MPO), which we constructed using the AutoMPO function of the ITensor.jl tensor networks toolkit~\cite{itensor}.
AutoMPO accepts a sum of products of operators and returns a chain of tensors whose contraction reproduces $\hat{\mathbb{W}}$, one tensor for each of the $L+1$ node voltages $v_0, \hdots, v_L$.
Because $\hat{\mathbb{W}}$ generates infinitesimal time evolution, it is a local operator.
In a tiny moment of time only one elementary hop occurs, and that hop changes the voltage at a single node at a rate set by that node and its neighbor.
Eq.~\eqref{eq:rateoperator} therefore carries nothing but nearest-neighbor terms, and a sum of such terms compiles into an MPO exactly.
Its bond index is a bookkeeping index rather than a physical one, counting at each cut the operator terms that must be carried from one side to the other.
The width of that index is therefore fixed by the structure of $\hat{\mathbb{W}}$, not by any truncation, and for Eq.~\eqref{eq:rateoperator} it never exceeds 6.
The operator's bond dimension is not a convergence parameter or a source of approximation.

The resulting MPO is translationally invariant in the bulk, since each unit contributes the same operators to Eq.~\eqref{eq:rateoperator} with only its site label shifted.
Only the two ends break the pattern, where nodes $0$ and $L$ interact with a single bulk unit rather than two.
That uniformity is a convenience, not a requirement.
Should the units differ\textemdash as fabrication variability or degradation can leave them~\cite{Asenov03,Mukhopadhyay05,Mahmoodi05,Maghsoudloo24}\textemdash the construction takes site-dependent rate operators and returns an MPO whose tensors vary from site to site.

\subsection{Basis Function Expansion}
\label{sec:basis_funcs}
Correlations between units affect the slow and fast dynamics of the chain unequally.
The inter-unit coupling reshapes the slow, global flipping dynamics, but it should leave the fast, local relaxations looking much as they do in an isolated unit.
That observation suggests a numerical opportunity: could the slow eigenvectors of a single CMOS unit seed a compact basis for the slow dynamics of the coupled chain?
To find out, we construct a reduced basis a priori from the slowest eigenvectors of the exactly solvable single-unit generator.
A successful reduction shrinks the physical dimension at every site, and with it the cost of every DMRG sweep.
Variational optimization can shape the ansatz parametrization itself, tuning it to capture rare events and large deviations~\cite{Banuls19,HelmsChan20,Casert21,Rose21}.
Our basis is fixed before any optimization begins, and DMRG subsequently converges only the MPS coefficients within it.
In other words, the basis sets what each site of the chain is allowed to look like, and DMRG converges only the tensors that link those local states together.
Other fixed-basis expansions, built from predefined weighting functions~\cite{Huang20,Cai24,Takahashi23} and from orthogonal polynomials~\cite{Ohkubo12}, share that premise, but they write their basis down analytically.
Ours is extracted numerically from the slow modes of the specific circuit at hand.

Solving $\mathbb{W}_{L=1}$ by exact diagonalization yields a set of right eigenvectors $\phi_i(v_0, v_1)$ defined over the two-dimensional voltage space of a single unit, ordered by the magnitude of the real part of their eigenvalues so that $\phi_0 = P_{\rm ss}^{(L=1)}$ is the single-unit steady state and increasing $i$ indexes progressively faster relaxation modes.
We retain only the $d$ slowest of these, reducing the per-node degrees of freedom from the full Fock dimension $D = 65$ to $d$; the truncation is the source of the computational savings, and $d$ is treated as a convergence parameter to be adjusted alongside the MPS bond dimension.
Each retained eigenvector is collapsed to a single-node function by marginalizing over one voltage index, exploiting the symmetry of the single-unit steady-state distribution,
\begin{equation}
    \varphi_i(v) = \sum_{v'} \phi_i(v', v),
    \label{eq:basis_marg}
\end{equation}
yielding $d$ functions of a single node's voltage.
Because these marginals are not in general orthogonal, we orthonormalize them by a QR decomposition to obtain an orthonormal set $\{\varphi_i(v)\}_{i=0}^{d-1}$, assembled as the rows of a rectangular map $U$ with $UU^\dagger = \mathbb{I}_d$.
These vectors collectively encode the slowest modes of the single-unit dynamics and, at large $V_{\rm dd}$, already capture the dominant slow structure of the coupled chain.

\begin{figure}
\includegraphics[width=\columnwidth]{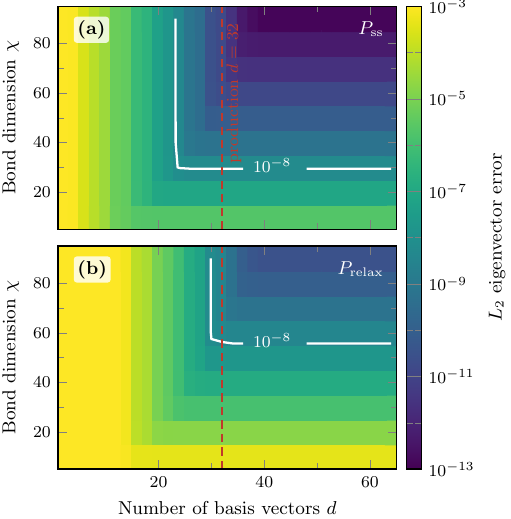}
\caption{\label{fig:KLdiv} Basis-set and bond-dimension convergence of the reduced-basis eigenvectors, as a function of the number of retained single-unit basis vectors $d$ (horizontal axis) and the MPS bond dimension $\chi$ (vertical axis) for $L=3$ units.
For each eigenvector the error is the $L_2$ norm $\|\bar{P}-P\|_2$ of its deviation from a reference converged with a sparse solver (GMRES). (a,~top)~Steady-state distribution $P_{\rm ss}$; (b,~bottom)~first relaxation mode $P_{\rm relax}$.
The white curve marks the $10^{-8}$ accuracy level and the dashed line the production basis cutoff $d=32$; both eigenvectors are converged well below $10^{-8}$ at the production settings ($d=32$, $\chi\leq100$).}
\end{figure}

We then project the rate operator onto this reduced basis at each site, $\hat{\mathbb{W}} \to U\hat{\mathbb{W}}U^\dagger$ applied site-by-site to the MPO, as illustrated in Fig.~\ref{fig:fig2}; the state MPS is mapped into the same basis by $U$ and back by $U^\dagger$.
This is a Galerkin projection onto the retained single-unit modes rather than a similarity transformation; it does not preserve the full spectrum, but by design it retains the slow modes that govern both $\ket{P_{\rm ss}}$ and $\ket{P_{\rm relax}}$.
How much of the basis can be discarded is a question we can settle directly at $L=3$.
That chain has $65^4\approx1.8\times10^7$ microstates, few enough for a sparse solver, so an independent reference is available there as it cannot be at $L=7$.
We computed the steady state and the first relaxation mode with GMRES and, separately, with full-basis, full-bond-dimension DMRG.
The two agree to machine precision, which establishes the reference; measuring the reduced basis against that reference then isolates the error introduced by truncation alone.
Figure~\ref{fig:KLdiv} shows that error, $\|\bar{P}-P\|_2$, across the plane of retained basis functions $d$ and bond dimension $\chi$.
Retaining $d = 32$ of the $D = 65$ Fock states, half the physical dimension, reproduces both eigenvectors to better than $10^{-8}$, and the production calculations use that cutoff.

\subsection{Numerical DMRG Convergence}
\label{sec:DMRGconvg}

Constructing the operator as an MPO cost us little.
Solving for its eigenspectrum is the hard part.
A single hop couples neighboring nodes, but $P_{\rm ss}(\mathbf{v})$ and $P_{\rm relax}(\mathbf{v})$ are shaped by many hops in succession, so the correlations they carry can reach across the chain.
Two things therefore stand between the MPO and the eigenvectors we want.
The solver must cope with a generator that is not Hermitian, and the truncations must be generous enough to hold those correlations.

AutoMPO is most commonly applied to Hermitian Hamiltonians, but it operates at the level of operator algebra to just as naturally generate non-Hermitian operators such as $\hat{\mathbb{W}}$.
Handed a sum of products of local operators, AutoMPO tracks which terms must be carried across each bond and assembles the matrices that reproduce the sum.
The bookkeeping manipulates terms, never spectra, and it never asks whether an operator equals its adjoint.
Though the non-Hermiticity of the Markov generator does not matter when building the MPO, the eigensolver called at each step of a sweep must be a non-Hermitian Krylov method, Arnoldi rather than Lanczos.

For a Hermitian operator each local update lowers a Rayleigh quotient, so the sweeps inherit a variational guarantee, and for a Markov generator that monotonicity is gone.
The identity of the fixed point survives.
By the Perron--Frobenius theorem the steady state is the leading eigenvector reached by the infinite-time propagation $\lim_{t\to\infty}e^{\hat{\mathbb{W}}t}\ket{p(0)}$, i.e.\ by the power method, and that infinite-time evolution is equivalent to an unbounded sequence of small-timestep time-dependent variational principle (TDVP) steps whose fixed point the DMRG sweeps reproduce~\cite{Zima25}.
The eigenvector $\ket{P_{\rm ss}}$ is therefore stationary under the sweeps even though $\hat{\mathbb{W}}$ is non-Hermitian.
The argument does not guarantee the optimization itself, since truncated Krylov subspaces and unfavorable initializations can still stall the sweeps, which is one motivation for the warm-start protocol described below.

Whether those truncations are generous enough for chains of $L\leq7$ is a numerical question, and one a Hermitian problem would face equally.
Figure~\ref{fig:KLdiv} varies both at once at $L=3$, the longest chain for which the GMRES reference of Sec.~\ref{sec:basis_funcs} can be computed.
Below a threshold bond dimension the error is insensitive to $d$ and set by $\chi$ alone, while above it the error is insensitive to $\chi$ and set by the basis alone.
The two eigenvectors are not equally demanding.
At the production basis $d=32$, an $L_2$ error of $10^{-8}$ requires $\chi=30$ for $P_{\rm ss}$ but $\chi=60$ for $P_{\rm relax}$, the slow mode being the less nearly factorizable of the two.
By $\chi\approx70$ neither improves further with $\chi$.
The residual $L_2$ errors, $2\times10^{-11}$ for $P_{\rm ss}$ and $3\times10^{-9}$ for $P_{\rm relax}$, are then set by the $d=32$ basis rather than by the bond dimension.
For the longer chains, where no such reference exists, we grow $\chi$ until the computed observables stop changing.

We perform sweeps of two-site DMRG to converge to an MPS approximation for $\ket{P_{\rm ss}}$, the top right eigenvector associated with the zero eigenvalue.
To compute the spectral gap $\mu_1$ and its associated eigenvector, $\ket{P_{\rm relax}}$, we must bias the iterative solver away from the steady-state~\cite{Stoudenmire12}.
We incorporate a penalty term into the rate operator of the form $\hat{\mathbb{W}}-\lambda\ket{P_{\rm ss}}\bra{P_{\rm ss}}$, where $\lambda$ is a sufficiently large constant to shift the steady-state eigenvalue below that of the next-slowest right eigenvector.
For a non-Hermitian operator, deflating with the right eigenvector $\bra{P_{\rm ss}}$ rather than the corresponding left eigenvector $\bra{\mathds{1}}$, the vector of all ones, is not in general guaranteed to preserve the target eigenvector; the generic prescription would instead use $\hat{\mathbb{W}}-\lambda\ket{P_{\rm ss}}\bra{\mathds{1}}$.
In our case, however, the voltage-inversion symmetry $\mathbf{v}\to-\mathbf{v}$ makes the simpler construction exact.
Under this inversion, $\ket{P_{\rm ss}}$ is symmetric while $\ket{P_{\rm relax}}$ is antisymmetric, so $\braket{P_{\rm ss}|P_{\rm relax}}=0$ and the penalty term annihilates the targeted eigenvector, leaving it an exact eigenvector of the modified operator (cf.\ the analogous symmetry argument in Ref.~\cite{Zima25}).
In practice, we use $\lambda=1.0$, which is generous for systems where $\mu_1\ll1$.

\begin{figure*}
  \includegraphics[width=\textwidth]{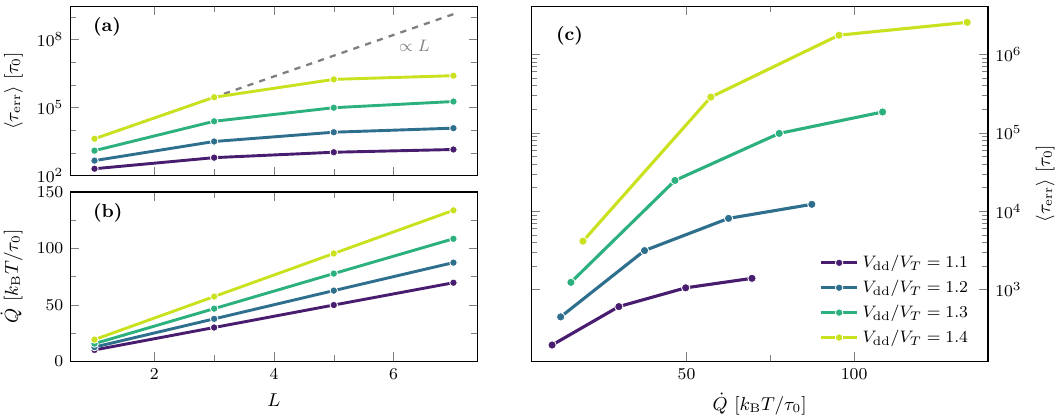}
\caption{\label{fig:fig3} Reliability and dissipation for CMOS chains. (a)~Mean bit-flip time as a function of chain length $L$ for several supply voltages $V_{\rm dd}$.
The growth of $\ln \langle\tau_{\rm err}\rangle$ with $L$ is sublinear, in contrast to the linear scaling expected from a reaction-diffusion analogy (dashed reference line). (b)~Steady-state dissipation as a function of $L$ for the same supply voltages, showing the expected linear-in-$L$ behavior. (c)~The resulting dissipation-reliability tradeoff.
Each marker corresponds to a fixed $(L, V_{\rm dd})$ pair; for any fixed dissipation budget (a vertical slice through the panel), the most reliable configuration is the one with minimum $L$ and maximum $V_{\rm dd}$.
Times are reported in units of the characteristic circuit timescale $\tau_0$, the average time to move a single electron through a transistor, and dissipation rates in units of $k_{\rm B}T/\tau_0$; both are defined in Sec.~\ref{sec:unit}.}
\end{figure*}

When converging both $\ket{P_{\rm ss}}$ and $\ket{P_{\rm relax}}$, we use a ``warm start'' protocol.
Beginning from an MPS initialized to give a uniform distribution over all voltage states, we perform DMRG sweeps with a small maximal bond dimension and with few retained basis functions ($d = 12$).
As the MPS converges, both cutoffs are increased (to $d = 32$), improving the MPS approximations of $\ket{P_{\rm ss}}$ and $\ket{P_{\rm relax}}$.
Because the sweeps are two-site, the bond dimension grows of its own accord as they proceed, up to whatever maximum is then in force.
We find we can capture the correlations in these distributions with a modest ($\leq 100$) bond dimension, computed in $5400$ sweeps across the warm-start stages for the most demanding case ($L=7$, $V_{\rm dd}/V_{\rm T}=1.4$) requiring approximately $260$ CPU-hours (1 core, 5 GB per run).
The numerical uncertainty in $\langle\tau_{\rm err}\rangle$ and $\dot{Q}$, estimated from the variation of $\mu_1$ over the final DMRG sweeps and from increasing the bond dimension, is smaller than the plot markers in Fig.~\ref{fig:fig3}.

One naturally asks what these eigenvectors look like, and for $L>1$ the question has no pictorial answer.
They are functions of all $L+1$ node voltages, so any rendering is a slice or a marginal rather than the object itself.
Their parity, however, is fixed by symmetry rather than by observation.
$\ket{P_{\rm ss}}$ is symmetric under $\mathbf{v}\to-\mathbf{v}$ and $\ket{P_{\rm relax}}$ antisymmetric, so the slow mode moves probability between the two macrostates at rate $\mu_1$.
Whether it does so by inverting every unit at once or by a localized route is not settled by the eigenvector, and we take up that question in Sec.~\ref{sec:results_L}.

Having computed the MPS for the steady-state distribution, $\dot{Q}$ is straightforward to extract.
We turn Eq.~\eqref{eq:deltaQ} into the diagonal operator
\begin{equation}
  \hat{\delta Q}_{\pm, i} \ket{\mathbf{v}} = e(-V_{\rm dd}+ \frac{v_e}{2} \pm v_i) \ket{\mathbf{v}},
\end{equation}
to get
\begin{equation}
  \dot{Q} = 2 \sum_{i = 0}^{L-1} \sum_{\alpha = \pm} \bra{\mathds{1}} \hat{\delta Q}_{\alpha, i+1}\left(\hat{t}^{\rm u}_{\alpha, i} - \hat{t}^{\rm d}_{\alpha, i} \hat{x}_{i+1}^{-\alpha}\right) \ket{P_{\rm ss}},
\end{equation}
Contracting with $\bra{\mathds{1}}$ performs the cheap sum over microstates in Eq.~\eqref{eq:dissipation}.
The all-ones vector is a product state, so the contraction never requires enumerating those microstates.
Note that $\alpha$ is a $\pm$ label, not a number.
The superscript in $\hat{x}_{i+1}^{-\alpha}$ therefore denotes the shift opposite to $\alpha$ rather than a power, just as the subscript on $\hat{\delta Q}_{\alpha, i+1}$ sets which sign of $v_{i+1}$ enters the heat.
The hop operators carry the index $i$ of the gate, whereas $\hat{\delta Q}_{\alpha, i+1}$ carries the index $i+1$ of the node it reads, matching the shift $\hat{x}_{i+1}^{-\alpha}$ that accompanies the same hop.
For compactness, we have double counted the dissipation through the top NOT gates since $\ket{P_{\rm ss}}$ is symmetric.

\section{Results: Reliability and Dissipation Scaling}
\label{sec:results}

We used the methodology of Sec.~\ref{sec:tn} to compute $\left<\tau_{\rm err}\right>$ and $\dot{Q}$ for chains of $L = 1$, $3$, $5$, and $7$ across the near-threshold window $V_{\rm dd}/V_T\in[1.1,1.4]$ of Sec.~\ref{sec:assumptions}, with the results collected in Fig.~\ref{fig:fig3}.
There are two distinct ways to make the bit last longer, increasing $V_{\rm dd}$ or increasing $L$.
Increasing either causes $\dot{Q}$ to grow linearly.
The two strategies are not interchangeable, however, because $\ln\left<\tau_{\rm err}\right>$ grows in proportion to $V_{\rm dd}^2$ but only sublinearly in $L$.

\subsection{Voltage Scaling of the Mean Bit-Flip Time}
\label{sec:results_vdd}

The dependence on $V_{\rm dd}$ is the simpler of the two.
A state-independent factor of $\exp(V_{\rm dd}/nV_T)$ can be factored out of both hopping rates in Eqs.~\eqref{eq:eq1} and~\eqref{eq:eq2}, and that factor merely rescales time.
Consequently, one component of $\ln\langle\tau_{\rm err}\rangle$ varies linearly in $V_{\rm dd}$ as an input to the model rather than a prediction of it.

\begin{figure}
\includegraphics[width=\columnwidth]{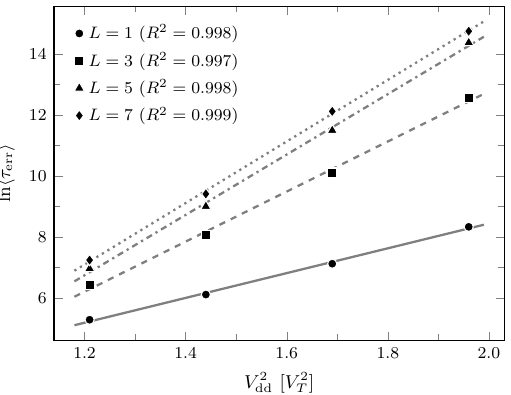}
\caption{\label{fig:vdd2} Mean bit-flip time versus squared supply voltage.
$\ln\langle\tau_{\rm err}\rangle$ is linear in $V_{\rm dd}^2$ for every chain length $L$ (markers, with linear fits), the signature of a quadratic Arrhenius barrier.
Derived analytically as a large-voltage asymptotic result for a single unit~\cite{Freitas22}, the quadratic scaling is seen here to persist for the coupled chains across the near-threshold window studied ($R^2 \geq 0.997$ for all $L$).}
\end{figure}

The remaining factor of $\exp(-V_{\rm dd}/V_T)$ in Eq.~\eqref{eq:eq2} is another matter.
That factor, which also appears in Eqs.~\eqref{eq:ldb} and~\eqref{eq:deltaQ}, sets the strength of the nonequilibrium driving, suppressing downhill hops and forcing the kinetics away from detailed balance.
How it shapes the barrier between the metastable states is far less obvious.
For a single CMOS unit, large-deviation theory yields a nonequilibrium quasipotential whose barrier grows quadratically with the supply voltage, $\Delta g \simeq \frac{2}{n+2}\,V_{\rm dd}^2/V_T$ to leading order at large $V_{\rm dd}$~\cite{Freitas22}.
An Arrhenius law therefore anticipates that $\ln\langle\tau_{\rm err}\rangle$ grows in proportion to $V_{\rm dd}^2$, and we find that the coupled chains follow the same law.
Linear fits of $\ln\langle\tau_{\rm err}\rangle$ against $V_{\rm dd}^2$, shown in Fig.~\ref{fig:vdd2}, give a coefficient of determination $R^2 \geq 0.997$ for every chain length studied.
That agreement was not guaranteed; the analytical barrier of Ref.~\cite{Freitas22} is a large-voltage ($V_{\rm dd} \gg V_T$) asymptotic result for a single unit, so its persistence for the coupled chains throughout our near-threshold window is an empirical finding.

\subsection{Length Scaling of the Mean Bit-Flip Time}
\label{sec:results_L}
No comparably rigorous analytical result guides the dependence of $\ln\langle\tau_{\rm err}\rangle$ on $L$.
We have an analogy instead.
In related work, we computed metastable transition times for a bistable reaction-diffusion model, a 1D chain of $L$ voxels, each housing a well-mixed bistable (Schl\"ogl) chemical reaction network with neighboring voxels exchanging particles by diffusion~\cite{Nicholson23}.
There, too, the log flip times grew in proportion to a barrier height, but the dependence on system size was qualitatively different, with $\ln\langle\tau_{\rm err}\rangle$ growing linearly in the number of voxels.
That linear growth came from the propagation, not from the nucleation.
The reaction-diffusion chain transitions by nucleating an interface at a boundary voxel and advancing it voxel by voxel, and the arrival of the interface tilts the next voxel's chemical bistability without removing it.
Every advance therefore remains a rare event in its own right, and the $L-1$ of them compound multiplicatively into a rate proportional to $p_{\rm int}^{L-1}$, i.e., $\ln\langle\tau_{\rm err}\rangle \propto L$~\cite{Nicholson23}.
By contrast, the CMOS chains exhibit \emph{sublinear} growth of $\ln\langle\tau_{\rm err}\rangle$ with $L$, as is apparent in Fig.~\ref{fig:fig3}(a), a nontrivial collective effect of the inter-unit correlations.

Why should two bistable chains scale so differently?
The scaling exponent itself narrows the possibilities.
A coherent inversion, in which all $L+1$ nodes cross together, must surmount a barrier that grows extensively with $L$, and that barrier would force $\ln\langle\tau_{\rm err}\rangle$ to grow at least linearly.
We observe sublinear growth, so the flip is not coherent.
The spectrum of $\hat{\mathbb{W}}$ excludes the opposite extreme.
Were the flip carried by independent localized switches, one available at each of the $L$ sites, one would expect a cluster of order $L$ nearly degenerate slow modes.
We used the penalty procedure of Sec.~\ref{sec:DMRGconvg} to compute the next dominant eigenvalue $\mu_2$ precisely to test for such a cluster, and we find none.
The gap remains large for all $L$, with $\mu_2/\mu_1 > 50$ for $L = 1$ and $\mu_2/\mu_1 > 100$ for $L \geq 3$.

That leaves a single coordinated interface.
A domain wall nucleates somewhere in the chain and then migrates, and it migrates at little additional cost.
The rare event is the nucleation alone, essentially independent of $L$, and the residual $L$ dependence enters as a weak, subextensive correction, precisely the sublinear growth of Fig.~\ref{fig:fig3}(a).
The reaction-diffusion chain cannot get that migration for free, and its activated advances are the origin of the linear growth there.
We caution that the scaling and the spectral gap constrain the mechanism without pinning it down.
The sharpest probe would be the committor for the flux between bit states~\cite{Strand24}, which requires the excited-state left as well as right eigenvectors together with a symmetry-aware construction of the reactive ensemble.
That machinery is the subject of future work.

\subsection{Dissipation-Reliability Tradeoff}
\label{sec:results_tradeoff}

Combining the two dependences gives the dissipation-reliability tradeoff of Fig.~\ref{fig:fig3}(c).
The advantage of increasing $V_{\rm dd}$ is particularly clear if the cost is the dissipation.
While the error time $\left<\tau_{\rm err}\right>$ is determined by rare events, $\dot{Q}$ is the typical steady-state dissipation, which is dominated by the dissipation through all transistors at the typical voltage.
The number of dissipating transistors grows linearly with $L$, and the typical voltage grows linearly with $V_{\rm dd}$.
Consequently, the dissipative cost $\dot{Q}$ grows linearly with both $L$ and $V_{\rm dd}$, as Fig.~\ref{fig:fig3}(b) confirms.

For a fixed dissipation budget, Fig.~\ref{fig:fig3} thus shows that the most robust bit is that with the smallest $L$.
However, it is not necessarily the case that dissipation minimization is always the goal.
Faced with the constraint to build a CMOS bit with small errors and with a maximal $V_{\rm dd}$, the results show that it can be efficacious to grow the chain.
Both scaling laws extrapolate to the device voltages of Sec.~\ref{sec:assumptions}, where the flip times are far too long to compute directly, so the same tradeoff governs the regime of practical interest.
Which strategy to choose there depends on whether the binding constraint is the dissipation budget or the supply voltage.

\section{Conclusion}
\label{sec:conclusion}
A CMOS bit can be made more reliable in two ways: by raising the supply voltage or by coupling units together into a chain.
Our calculations quantify both strategies and the dissipation each incurs.
Raising $V_{\rm dd}$ achieves equivalent reliability at lower dissipative cost, while lengthening the chain offers error suppression in the low-voltage regimes motivated by miniaturized and implantable electronics.
The length scaling proved the subtler of the two: $\ln\langle\tau_{\rm err}\rangle$ grows sublinearly in $L$, in contrast to the linear-in-$L$ scaling of analogous bistable reaction-diffusion networks~\cite{Nicholson23}, and together with the isolated slow relaxation mode, that scaling supports domain-wall nucleation, rather than a coherent flip of all nodes, as the mechanism of bit failure.

Those physical results required numerical access to rare events in enormous state spaces, which we obtained by introducing a tensor network approach for coupled stochastic electronic circuits.
Using DMRG with a judicious basis set expansion, we computed the top two right eigenvectors of a thermodynamically consistent circuit master equation; the first yields the steady-state dissipation rate, and the second yields the mean bit-flip time.
The approach preserves the discreteness of individual electron hops, and with it the local detailed balance that ties circuit energetics to thermal fluctuations, for state spaces exceeding $10^{14}$ microstates.

Several future directions follow naturally.
The committor calculation of Sec.~\ref{sec:results_L} will make the nucleation picture rigorous~\cite{Strand24,Chen23}.
The same framework accommodates chains with device variability~\cite{Asenov03,Mukhopadhyay05,Mahmoodi05,Maghsoudloo24}, in which heterogeneity shapes both the stability of the bit and the structure of the flip pathway.
Such heterogeneity enters through the parameters of the master equation, which need not be the same at every site.
Whether a mismatched chain nucleates its flip at its weakest unit is a question that our uniform chains cannot pose, and that committor calculation could settle it.
More complicated architectures are within reach as well~\cite{Ueda14,Wolpert20}, but the chain geometry itself is harder to relax.
Closing the chain into a ring introduces a loop into the tensor network, and that loop spoils the cheap sequential contraction of a matrix product state, so periodic boundaries remain the awkward case for DMRG.
Developments in methods for looped tensor networks~\cite{Midha25} could extend the calculations to rings and periodic boundary conditions.

Nothing about the approach is particular to CMOS.
It requires only a master equation on a high-dimensional discrete state space, and we anticipate that it will prove useful for any p-bit platform that can be modeled in that way~\cite{Bhatti17,Borders19,BP24,Elmitwalli23}, just as the two eigenvectors computed here offer a way to assess the proximity of CMOS architectures to fundamental limits on charging speed and computational modularity~\cite{Gao21,Wolpert20,Helms25,Chen26,Boyd18,Riechers20}.
Other stochastic electronic technologies~\cite{Rivnay13,Sun18,Jia21,Dowling21,Slipko22} should invite the same treatment, including those developed for thermodynamic computing algorithms~\cite{Aifer24,Melanson25,Whitelam26,Kim24,Coles23,Aifer24TLA,TCconference,Freitas26,Chowdhury23}, which harness the thermal noise that flips our bit rather than fighting it.
When the noise is doing the computing, the design objective inverts: one wants a short mean flip time at fixed dissipation rate rather than a long one.
Can the same pair of eigenvectors pick out the circuit designs that achieve it?
Even more ambitiously, similar calculations could illuminate robustness-dissipation tradeoffs in biochemical reaction networks~\cite{Zima25,Tkacik25,Ouldridge17,Nicholson25}, perhaps leveraging applications of modular circuit theory to chemical reaction networks~\cite{Avanzini23,Raux25}.

\bigskip
\begin{acknowledgments}
This material is based upon work supported by the National Science Foundation under Grant No.~2239867 (T.R.G.) and by the National Science Foundation Graduate Research Fellowship Program under Grant No.~DGE-2234667 (C.M.).
Any opinions, findings, and conclusions or recommendations expressed in this material are those of the authors and do not necessarily reflect the views of the National Science Foundation.
This research was supported in part through the computational resources and staff contributions provided for the Quest high performance computing facility at Northwestern University, which is jointly supported by the Office of the Provost, the Office for Research, and Northwestern University Information Technology.
\end{acknowledgments}

\bigskip
\noindent\textit{Data and Code Availability.}---The code used to construct the MPO, perform the DMRG calculations, and generate the reported observables is openly available~\cite{Murphy26code}.

\bibliography{refs}

\end{document}